\documentclass[a4paper]{jpconf}

\usepackage{amsfonts}
\usepackage{amssymb}
\usepackage{amsbsy}
\usepackage{amsmath}
\usepackage{graphicx}

 \newcommand{\invfb}{\,{\rm fb}^{-1}}

\newcommand{\gev}{{\hbox{GeV}}}

\newcommand{\mbc}{{M_{\textrm{bc}}}}
\newcommand{\deltae}{{\Delta E}}

\newcommand{\lint}{{L_{\textrm{int}}}}
\newcommand{\trho}{\theta_{\rho^+}}
\newcommand{\tdsst}{\theta_{\dsst}}
\newcommand{\crho}{\cos\trho}
\newcommand{\cdsst}{\cos\tdsst}
\newcommand{\ssqdsst}{\sin^2\tdsst}
\newcommand{\csqrho}{\cos^2\trho}
\newcommand{\csqdsst}{\cos^2\tdsst}
\newcommand{\ssqrho}{\sin^2\trho}
\newcommand{\dG}{\Delta\Gamma_s^{CP}}
\newcommand{\G}{\Gamma_s}
\newcommand{\rfphi}{R_{f/\phi}}

\newcommand{\BR}{{\mathcal B}}

\newcommand{\FourS}{\Upsilon(4S)}
 \newcommand{\FiveS}{\Upsilon(5S)}
 \renewcommand{\bs}{{B_s^0}}
\newcommand{\bsst}{{B_s^{\ast}}}
\newcommand{\barbsst}{{{\bar B_s}^{\ast}}}
\newcommand{\bsST}{{B_s^{(\ast)}}}
\newcommand{\ds}{{D_s^-}}
\newcommand{\dsst}{D_s^{\ast-}}

\newcommand{\KS}{{K_S^0}}

\newcommand{\jpsi}{{J\!/\!\psi}}
\newcommand{\fz}{f_0(980)}

\newcommand{\bsSTbsST}{{\bsST{\bar B_s}^{(\ast)}}}
\newcommand{\dsSTdsST}{D_s^{(\ast)+}D_s^{(\ast)-}}
\newcommand{\dsstdsst}{D_s^{\ast+}D_s^{\ast-}}
\newcommand{\dsstds}{D_s^{\ast\pm}D_s^{\mp}}
\newcommand{\dsds}{D_s^{+}D_s^{-}}

\newcommand{\bsdspi}{{\bs\to D_s^-\pi^+}}
\newcommand{\bsdsstpi}{{\bs\to D_s^{\ast-}\pi^+}}
\newcommand{\bsdsSTpi}{{\bs\to D_s^{(\ast)-}\pi^+}}
\newcommand{\bsdsk}{{\bs\to D_s^{\mp}K^{\pm}}}
\newcommand{\bsdsrho}{{\bs\to D_s^-\rho^+}}
\newcommand{\bsdsstrho}{{\bs\to D_s^{\ast-}\rho^+}}
\newcommand{\bsdsSTrho}{{\bs\to D_s^{(\ast)-}\rho^+}}
\newcommand{\bsjpsieta}{{\bs\to\jpsi\,\eta}}
\newcommand{\bsjpsietaP}{{\bs\to\jpsi\,\eta^{(')}}}
\newcommand{\bsjpsietap}{{\bs\to\jpsi\,\eta^{'}}}

\newcommand{\bsjpsifz}{{\bs\to\jpsi\,\fz}}
\newcommand{\bspipi}{{\bs\to\pi^+\pi^-}}
\newcommand{\bskpi}{{\bs\to K^-\pi^+}}
\newcommand{\bskk}{{\bs\to K^+K^-}}
\newcommand{\bskzkz}{{\bs\to K^0\bar K^0}}

\newcommand{\bfbstodsk}{(2.4^{+1.2}_{-1.0}({\textrm{stat.}})\pm0.3({\textrm{syst.}})\pm0.3(f_s))\times10^{-4}}

\newcommand{\fss}{{\left(90.1^{+3.8}_{-4.0}\pm0.2\right)\%}}

\newcommand{\fssp}{{90.1^{+3.8}_{-4.0}\pm0.2}}
\newcommand{\fsp}{{7.3^{+3.3}_{-3.0}\pm0.1}}
\newcommand{\ffp}{{2.6^{+2.6}_{-2.5}}}

\newcommand{\mbs}{{5364.4\pm1.3\pm0.7}}
\newcommand{\mbsst}{{5416.4\pm0.4\pm0.5}}

\renewcommand{\fl}{{1.05^{+0.08}_{-0.10}{}^{+0.03}_{-0.04}}}

\newcommand{\bfbstojpsieta}{{(3.32\pm0.87({\rm stat.}){}^{+0.32}_{-0.28}({\rm syst.})\pm0.42(f_s))\times10^{-4}}}
\newcommand{\bfbstojpsietaprime}{{(3.1\pm1.2({\rm stat.})^{+0.5}_{-0.6}({\rm syst.})\pm0.4(f_s))\times 10^{-4}}}

\newcommand{\bfbstokk}{{(3.8{}^{+1.0}_{-0.9}({\rm stat.})\pm0.5({\rm syst.})\pm0.5(f_s))\times 10^{-5}}}
\newcommand{\bfbstokpi}{{2.6\times10^{-5}}} %limit
\newcommand{\bfbstopipi}{{1.2\times10^{-5}}} %limit
\newcommand{\bfbstokzkz}{{6.6\times 10^{-5}}} %limit

\newcommand{\bfbstodsstdsst}{{(3.1^{+1.2}_{-1.0}({\rm stat.})\pm0.8({\rm syst.}))\%}}
\newcommand{\bfbstodsSTdsST}{{(6.9^{+1.5}_{-1.3}({\rm stat.})\pm1.9({\rm syst.}))\%}}
\newcommand{\bfbstodsstds}{{(2.8^{+0.8}_{-0.7}({\rm stat.})\pm0.7({\rm syst.}))\%}}
\newcommand{\bfbstodsds}{{(1.0^{+0.4}_{-0.3}({\rm stat.})^{+0.3}_{-0.2}({\rm syst.}))\%}}
\newcommand{\dGoG}{{(14.7^{+3.6}_{-3.0}({\rm stat.}){}^{+4.4}_{-4.2}({\rm syst.}))\times10^{-2}}}

\newcommand{\bfbstojpsifz}{{1.63\times10^{-4}\textrm{~(at 90\% C.L.)}}} %limit

\begin{document}

\title{$\pmb{B_s^0}$ Decays at Belle}

\author{Remi Louvot}
\address{  (On behalf of the Belle collaboration)\\
  \'Ecole Polytechnique F\'ed\'erale de Lausanne~(EPFL), Lausanne, Switzerland}
\ead{remi.louvot@epfl.ch}

\begin{abstract}
  The large data sample recorded with the Belle detector at the
  $\FiveS$ energy provides a unique opportunity to study the poorly-known $\bs$ meson.
  Several analyses, made with a data sample representing an integrated luminosity of 23.6~$\invfb$, are presented.
  We report the study of the large-signal $\bs\to D_s^{(\ast)-}h^+$ ($h^+=\pi^+,\rho^+$) decays.
  In addition, several results on $\bs$ decays related to $CP$-violation studies are described.
  Beside the non-flavor specific $\bsdsk$ decay, the following $CP$-eigenstate decays are studied:
  $\bsjpsietaP$, $\bsjpsifz$, the charmless $\bs\to h\bar h$ ($h=\pi^+,K^+,\KS$) 
  and the three $\bs\to\dsSTdsST$ modes from which $\dG/\G$ is extracted.
  
  Talk presented at the \emph{Symposium on Prospects in the Physics of Discrete Symmetries}
  (Rome, Italy, 6-11 December 2010).
  LPHE Note 2011-01 (dated \today).
\end{abstract}

\section{Introduction}
The Belle experiment \cite{NIMA_479_117}, located at the interaction point of
the KEK $B$ factory \cite{NIMA_499_1},
was designed for the study of $B$ mesons\footnote{The notation ``$B$'' refers either to a $B^0$ or a $B^+$.
Moreover, charge-conjugated states are implied everywhere.}
produced in $e^+e^-$ annihilation at the $\FourS$ resonance ($\sqrt s\approx10.58$ GeV).
After having recorded an unprecedented sample of $\sim550$ millions of $B\bar B$ pairs,
the Belle collaboration started to record collisions at higher energies,
opening the possibility to study the $\bs$ meson.
A significant theoretical effort on the $\bs$ potential for studying $CP$ violation in the Standard Model (SM) 
and beyond has taken place from the 90's \cite{PRD_52_3048}. %,PRD_63_114015}.
The non-flavor specific decays such as $\bsdsk$ are expected to be insensitive to new physics (NP);
they can be used to measure the CKM angle $\gamma$ \cite{ZPC_54_653} %,NPB_671_459}
and help to resolve the ambiguity in the width difference in the $\bs-\bar\bs$ system,
$\Delta\Gamma_s$ \cite{PRD_77_054010}.
The penguin-dominated decays $\bs\to K^+K^-$ are expected to be sensitive to NP \cite{PRD_70_031502} 
and the CKM angles $\beta$ and $\gamma$ can be extracted by comparing its branching fraction (BF) with that of $B^0\to\pi^+\pi^-$ \cite{PLB_459_306}.
Finally, the decays involving a $b\to c\bar cs$ transition
(the ``golden mode'' $\bs\to\jpsi\phi$, $\bsjpsietaP$, $\bsjpsifz$, $\bs\to\dsSTdsST$, etc.)
are probably the most promising,
as the $CP$-violation effects in these modes are expected to be tiny in the SM \cite{PLB_475_111}.
A particular attention will be given to the $\bs\to\dsSTdsST$ modes
because the width difference $\Delta\Gamma_s^{CP}$ can be extracted
from their total BF \cite{PLB_316_567}.

Up to now, a data sample of integrated luminosity of $\lint=(23.6\pm0.3)\invfb$ (out of a total of $120~\invfb$)
has been analyzed at the energy of the $\FiveS$ resonance
($\sqrt s\approx10.87~\gev$).
Since the $\FiveS$ resonance is above the $\bs\bar\bs$ threshold, it was naturally
expected that the $\bs$ meson could be studied with $\FiveS$ data as well as
the $B$ mesons are with $\FourS$ data.
The main advantage with respect to the hadron colliders is
the better prospects of absolute BF measurements.
However, the abundance of $\bs$ mesons in $\FiveS$ hadronic events has to be
precisely determined.
Above the $e^+e^-\to u\bar u, d\bar d, s\bar s, c\bar c$ continuum events,
the $e^+e^-\to b\bar b$ process can produce different kinds of final states involving
a pair of non-strange $B$ mesons \cite{PRD_81_112003} ($B^{\ast}\bar B^{\ast}$,
$B^{\ast}\bar B$, $B\bar B$, $B^{\ast}\bar B^{\ast}\pi$, $B^{\ast}\bar B\pi$,
$B\bar B\pi$, $B\bar B\pi\pi$ and $B\bar B\gamma$),
a pair of $\bs$ mesons ($\bsst\barbsst$, $\bsst\bar\bs$ and $\bs\bar\bs$),
or final states involving a lighter bottomonium resonance below the open-beauty
threshold \cite{PRL_100_112001}.
The $B^{\ast}$ and $\bsst$ mesons always decay by emission of a photon.
The total $e^+e^-\to b\bar b$ cross section at the $\FiveS$ energy was measured
to be $\sigma_{b\bar b}=(302\pm14)$~pb
\cite{PRL_98_052001,PRD_75_012002} and the fraction of $\bs$ events
to be\footnote{The BF values presented here are calculated with $f_s=(19.3\pm2.9)\%$ \cite{PDG10}.
The BFs of $\bsdspi$, $\bsdsk$ and those in Sections \ref{sec:bshh} and \ref{sec:jpsi} 
are calculated with $f_s=(19.5^{+3.0}_{-2.3})\%$, also provided in Ref.~\cite{PDG10}.}
$f_s=\sigma(e^+e^-\to\bsSTbsST)/\sigma_{b\bar b}=(20.2\pm3.6)$~\% \cite{hepex_1010_1589}.
The dominant $\bs$ production mode, $b\bar b\to\bsst\barbsst$,
represents a fraction $f_{\bsst\barbsst}=\fss$ of the $b\bar b\to\bsSTbsST$ events,
as measured with $\bsdspi$ events (see next section and Ref.~\cite{PRL_102_021801}).

For all the exclusive modes presented here, the $\bs$ candidates are fully reconstructed
from the final-state particles.
From the reconstructed four-momentum in the CM, $(E_{\bs}^{\ast},\pmb{p}_{\bs}^{\ast})$,
two variables are formed:
the energy difference $\deltae=E_{\bs}^{\ast}-\sqrt s/2$ and the
beam-constrained mass $\mbc=\sqrt{s/4-\pmb{p}_{\bs}^{\ast2}}$.
The signal coming from the dominant $e^+e^-\to\bsst\barbsst$ production mode
is extracted from a two-dimensional fit performed on the distribution of these
two variables.
The corresponding BF is then extracted using the total efficiency (including sub-decay BFs)
determined with Monte-Carlo (MC) simulations, $\sum\varepsilon\BR$,
and the number of $\bs$ mesons produced via the $e^+e^-\to\bsst\barbsst$ process,
$N_{\bs}=2\times\lint\times\sigma_{b\bar b}\times f_s\times f_{\bsst\bar\bsst}\sim 2.5\times10^6$.

\section{Dominant CKM-favored $\pmb{\bs}$ Decays}
We report the measurement of exclusive $\bs\to D_s^{(\ast)-}h^+$ ($h^+=\pi^+$ or $\rho^+$) decays \cite{PRL_102_021801}.
These modes are expected to produce an abundant signal
because of their relatively large predicted BFs \cite{PLB_318_549}
and their clean signatures: four charged tracks and up to two photons.
The leading amplitude for the four $\bsdsSTpi$ and $\bsdsSTrho$ modes
is a $b\to c$ tree diagram of order $\lambda^2$
(in the Wolfenstein parametrization~\cite{PRL_51_1945}
of the CKM quark-mixing matrix~\cite{PRL_10_531}) with a spectator $s$ quark.
Besides being interesting in their own right, such measurements, if precise enough, 
can be of high importance for the current and forthcoming hadron collider experiments. 
It was for example pointed out~\cite{hepex_0912_4179} that the search 
for the very rare decay $\bs\to\mu^+\mu^-$ will be 
systematically limited by the poor knowledge of $\bs$ production, in case 
NP will enhance the decay probability by no more than a factor 3 
above the SM expectation. 

\begin{figure}[!htb]
  \centering
  \begin{minipage}{0.6\linewidth}
    \centering
    \includegraphics[width=\linewidth]{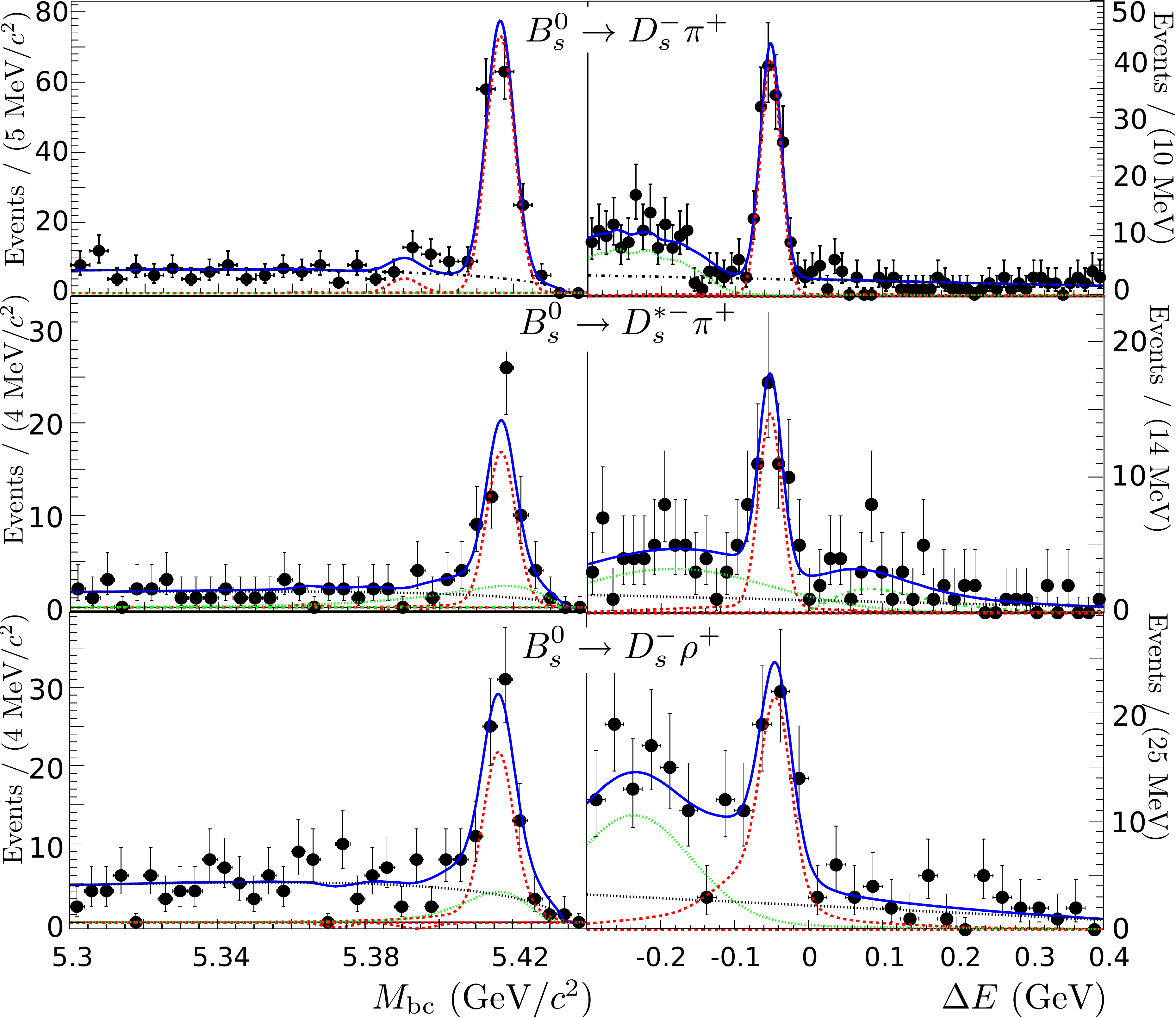}
  \end{minipage}
  ~~
  \begin{minipage}{0.35\linewidth}
    \caption{\label{fig2} Left: $\mbc$ distributions for the $\bsdspi$ (top) $\bsdsstpi$ (middle) and $\bsdsrho$ (bottom) candidates
      with $\deltae$ restricted to the $\bsst\barbsst$ signal region.
      Right: $\deltae$ distributions with $\mbc$ restricted to the $\bsst\barbsst$ signal region.
      The black- (green-) dotted curve represents the continuum (peaking) background,
      while the red-dashed curves are the signal shapes.
      The larger one is the signal in the $\bsst\barbsst$ kinematic region and the two others,
      which are very close to 0, are the signals in the two other $\bs$ production modes ($\bsst\bar\bs$ and $\bs\bar\bs$).  }
  \end{minipage}
\end{figure}

In addition, polarization measurements of $B$ decays have become of high interest since the observation
of a surprisingly large transverse polarization in $B\to\phi K^{\ast}$ decays by BaBar \cite{PRL_91_171802} and Belle \cite{PRL_91_201801}.
The relative strengths of the longitudinal and transverse states can be
measured with an angular analysis of the decay products.
In the helicity basis, the expected $\bsdsstrho$ differential decay width is proportional to 
$\frac{{\rm d}^2\Gamma(\bsdsstrho)}{{\rm d}\cdsst {\rm d}\crho}\propto 4f_L\ssqdsst\csqrho+(1-f_L)(1+\csqdsst)\ssqrho$,
where $f_L$ is the longitudinal polarization fraction,
and $\tdsst$ ($\trho$)  is the helicity angle of the $D_s^{\ast-}$ ($\rho^+$)
defined as the supplement of the angle between the $\bs$ and the $\ds$ ($\pi^+$) momenta
in the $\dsst$ ($\rho^+$) frame.

\begin{figure}[!htb]
  \centering
  \begin{minipage}{0.6\linewidth}
    \centering
    \includegraphics[width=\linewidth]{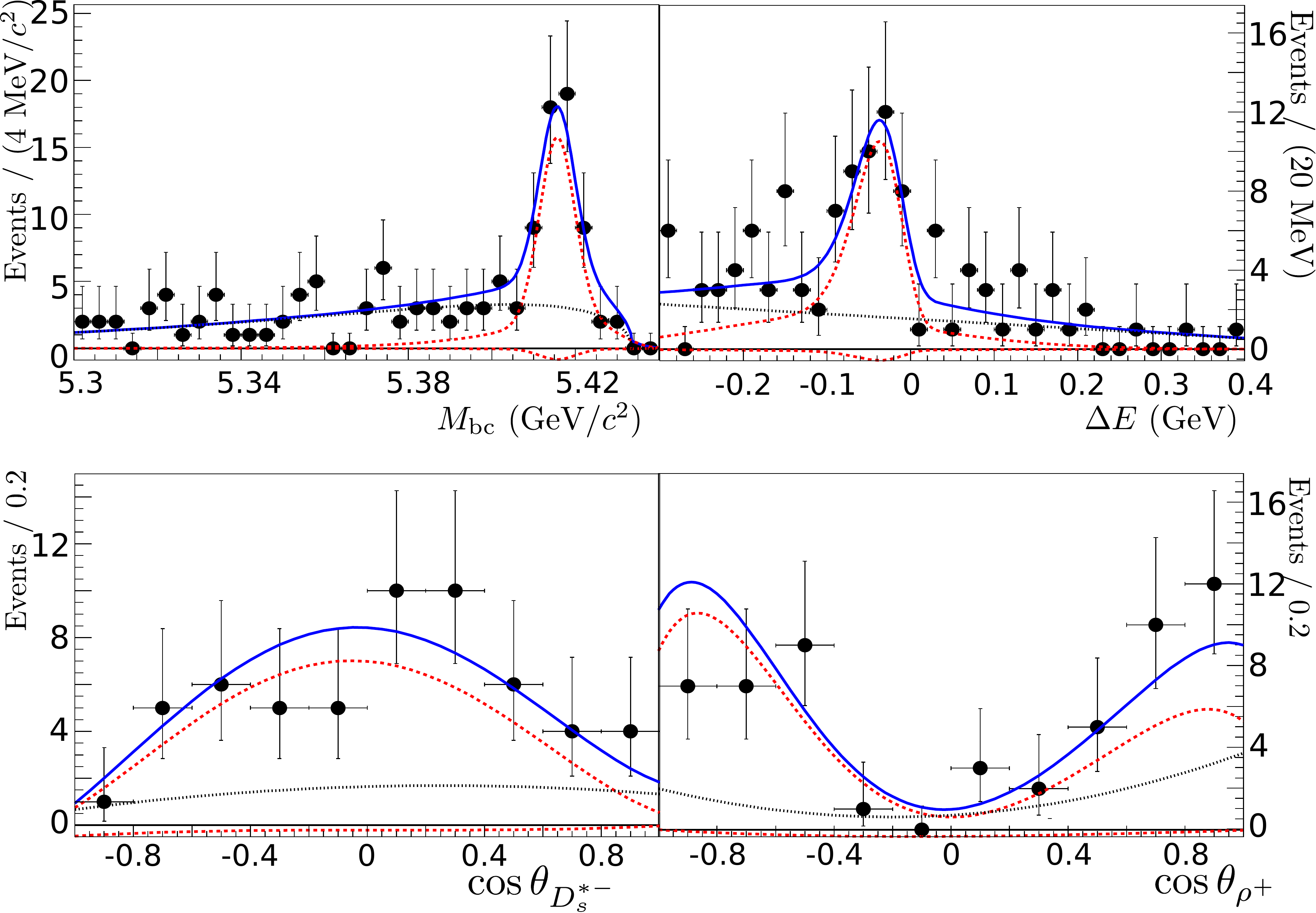}
  \end{minipage}
  \begin{minipage}{0.35\linewidth}
    \caption{\label{fig3}     
      Fit of the $\bsdsstrho$ candidates.
      Top: $\mbc$ and $\deltae$ distributions, similarly to Fig.~1. 
      Bottom: helicity distributions of the $D_s^{\ast -}$ (left) and $\rho^+$ (right) with $\mbc$
      and $\deltae$ restricted to the $\bsst\barbsst$ kinematic region.
      The black-dotted curve represents the background, while the two red-dashed curves are the signal.
      The large (small) signal shape corresponds to the longitudinal (transverse) component.
    } 
  \end{minipage}
\end{figure}

The $\ds$ mesons are reconstructed via three modes :
$\ds\to\phi(\to K^+K^-)\pi^-$, $\ds\to K^{\ast0}(\to K^+\pi^-)K^-$ and
$\ds\to\KS(\to\pi^+\pi^-)K^-$.
Based on the ratio of the second and the zeroth Fox-Wolfram moments \cite{PRL_41_1581}, $R_2$,
the continuum events are efficiently
rejected by taking advantage of the difference between their event geometry (jet like, high $R_2$) and
the signal event shape (spherical, low $R_2$).
The $\bsdspi$ and $\bsdsstpi$ ($\bsdsrho$ and $\bsdsstrho$) candidates with $R_2$ smaller than 0.5 (0.35) are kept for further analysis.
A best candidate selection, based on the intermediate-particle reconstructed masses, is then implemented in order to keep only one $\bs$ candidate per event per mode.
The $\mbc$ and $\deltae$ distributions of the selected $\bs$ candidates for the three $D_s^-$
modes are shown in Figs.~\ref{fig2} and \ref{fig3}, where the various components of the probability density function (PDF) used for the fit are described.
The $\bsdsstrho$ candidates are observed with two additional variables, $\cdsst$ and $\crho$,
which are the cosines of the helicity angles defined above.
They are needed for the measurement of $f_L$.

Table~\ref{summary} presents a summary of the numerical results obtained for the $\bsdsSTpi$ and $\bs\to$ $D_s^{(\ast)-}\rho^+$ modes.
The different sources of systematic uncertainties affecting the measurements are identified and quoted as a second error.
Our results on the $\bs$ decays are consistent with theoretical predictions \cite{PLB_318_549} and with existing measurements (Table~\ref{summary}).

\begin{table}[!ht]
  \centering
  \renewcommand{\arraystretch}{1.3}
  \caption{\label{summary}
    Summary of the results for the four $\bsdsSTpi$ and $\bsdsSTrho$ modes \cite{PRL_102_021801}.
    Top: signal yields in the $\bsst\barbsst$ production mode ($N_{\bsst\barbsst}$), 
    significances ($S$) including systematics,
    total signal efficiencies ($\varepsilon$) including all sub-decay BFs, and
    BFs ($\BR$), where the uncertainty due to $f_s$ (third error) is separated from the others systematics (second error).
    The first error represents the statistical uncertainties.
    Bottom: other measurements, obtained with the $\bsdspi$ and $\bsdsstrho$ analyses.
  }
  
  \begin{tabular}{lc@{\hspace{0.4cm}}c@{\hspace{0.4cm}}c@{\hspace{0.4cm}}c}
    \br
    Mode         & $N_{\bsst\barbsst}$    & $S$          & $\varepsilon$ ($10^{-3}$)      & $\BR$ ($10^{-3}$)              \\
    \mr
    $\bsdspi$    & $145^{+14}_{-13}$      & $21\sigma$   & $15.8$                         & $3.7^{+0.4}_{-0.3}\pm0.4\pm0.5$  \\
    $\bsdsstpi$  & $53.4^{+10.3}_{-9.4}$  & $7.1\sigma$  & $9.13$                         & $2.4^{+0.5}_{-0.4}\pm0.3\pm0.4$  \\
    $\bsdsrho$   & $92.2^{+14.2}_{-13.2}$ & $8.2\sigma$  & $4.40$                         & $8.5^{+1.3}_{-1.2}\pm1.1\pm1.3$  \\
    $\bsdsstrho$ & $77.8^{+14.5}_{-13.4}$ & $7.4\sigma$  & $2.67$                         & $11.9^{+2.2}_{-2.0}\pm1.7\pm1.8$ \\
  \end{tabular}
  
  \begin{tabular}{lc}
    \mr
    Observable          &Our measurement \\
    \mr
    Masses (MeV/$c^2$)& $m(\bs)=\mbs$ ;  $m(\bsst)=\mbsst$  \\
    Prod. mode (\%)& $f_{\bsst\barbsst}=\fssp$ ;    $f_{\bsst\bar\bs}=\fsp$  ;     $f_{\bs\bar\bs}=\ffp$     \\
    $f_L(\bsdsstrho)$   & $\fl$     \\
    \br
  \end{tabular}
\end{table}

\section{Evidence for $\pmb{\bsdsk}$}
\begin{figure}[!htb]
  \centering
  \begin{minipage}{0.27\linewidth}
    \centering
    \includegraphics[height=3.5cm,width=\linewidth]{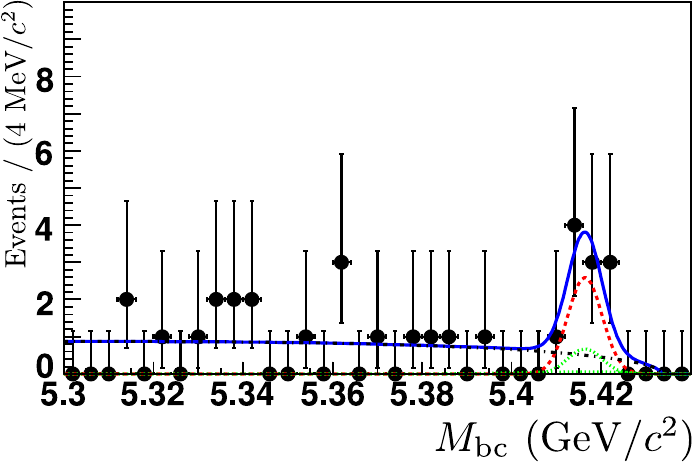}
  \end{minipage}~
  \begin{minipage}{0.27\linewidth}
    \centering
    \includegraphics[height=3.5cm,width=\linewidth]{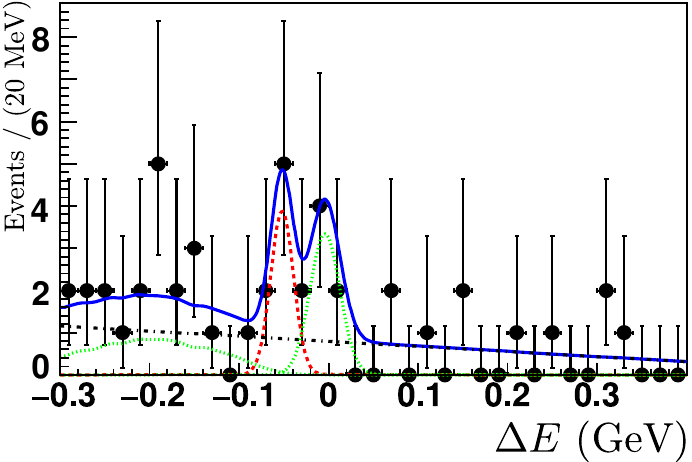}
  \end{minipage}~~~~
  \begin{minipage}{0.42\linewidth}
    \caption{\label{dsk}Distributions, similarly to Fig.~1, of the $\bsdsk$ candidates and the fitted PDF as solid blue curve.
      The red, black and green dotted curves represent the signal, the continuum and the $\bsdspi$ component of the PDF, respectively.}
  \end{minipage}
\end{figure}

We searched for the Cabibbo-suppressed counterpart of $\bsdspi$, $\bsdsk$, which is not flavour specific \cite{PRL_102_021801}.
The analysis is performed in the same way as that of $\bsdspi$, by replacing the $\pi^+$ by a $K^+$ candidate.
The fit also includes the $\bsdspi$ contamination (when the pion is misidentified as a kaon).
A 3.5$\sigma$ evidence with $6.7^{+3.4}_{-2.7}$ events is obtained in the $\bsst\barbsst$ signal region (Fig.~\ref{dsk}),
leading to the branching fraction $\BR(\bsdsk)=\bfbstodsk$, in agreement with the CDF result \cite{PRL_103_191802}.

\section{Observation of $\pmb{\bskk}$ and Searches for $\pmb{\bspipi}$, $\pmb{\bskpi}$ and $\pmb{\bs\to\KS\KS}$}\label{sec:bshh}
We present our results for the $\bskk$, $\bskpi$, $\bspipi$ and $\bs\to\KS\KS$ charmless decays \cite{PRD_82_072007}.
The charged pion and kaon candidates are selected using charged tracks and identified with energy deposition, momentum and time-of-flight measurements.
The $\KS$ candidates are reconstructed via the $\KS\to\pi^+\pi^-$ decay.
A likelihood based on 16 modified Fox-Wolfram moments \cite{PRL_91_261801} is implemented to reduce the continuum, which is the main source of background.

\begin{figure}[!tb]
  \centering
  \begin{minipage}{0.3\linewidth}
    \centering
    \includegraphics[height=4cm,width=\linewidth]{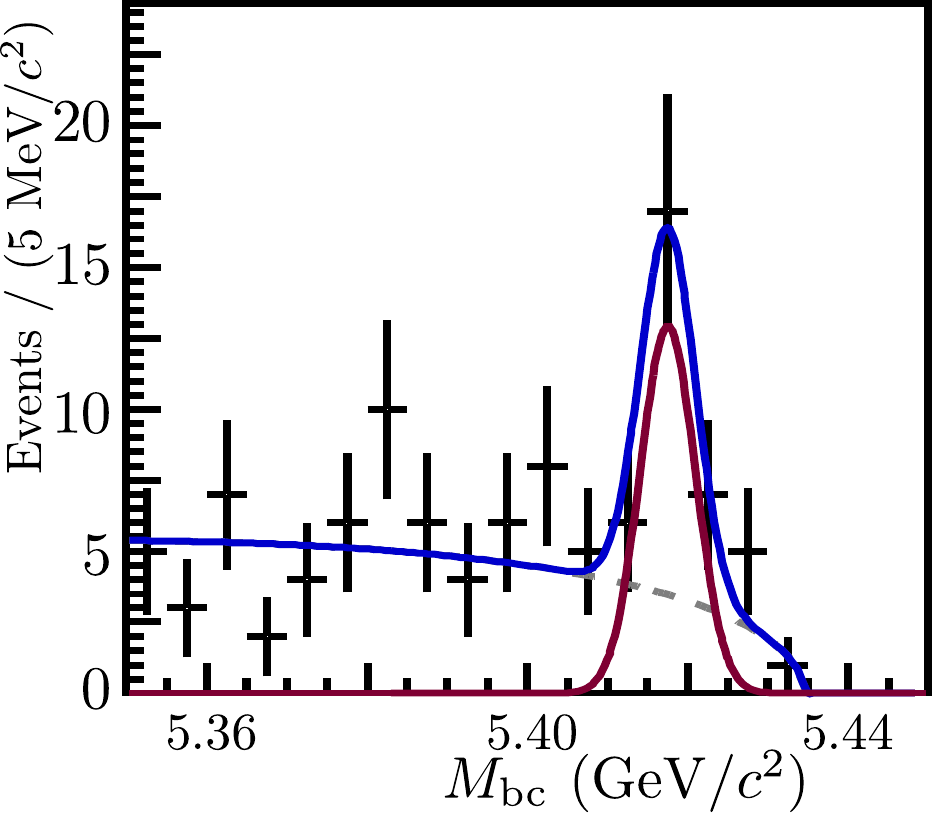}
  \end{minipage}~
  \begin{minipage}{0.3\linewidth}
    \centering
    \includegraphics[height=4cm,width=\linewidth]{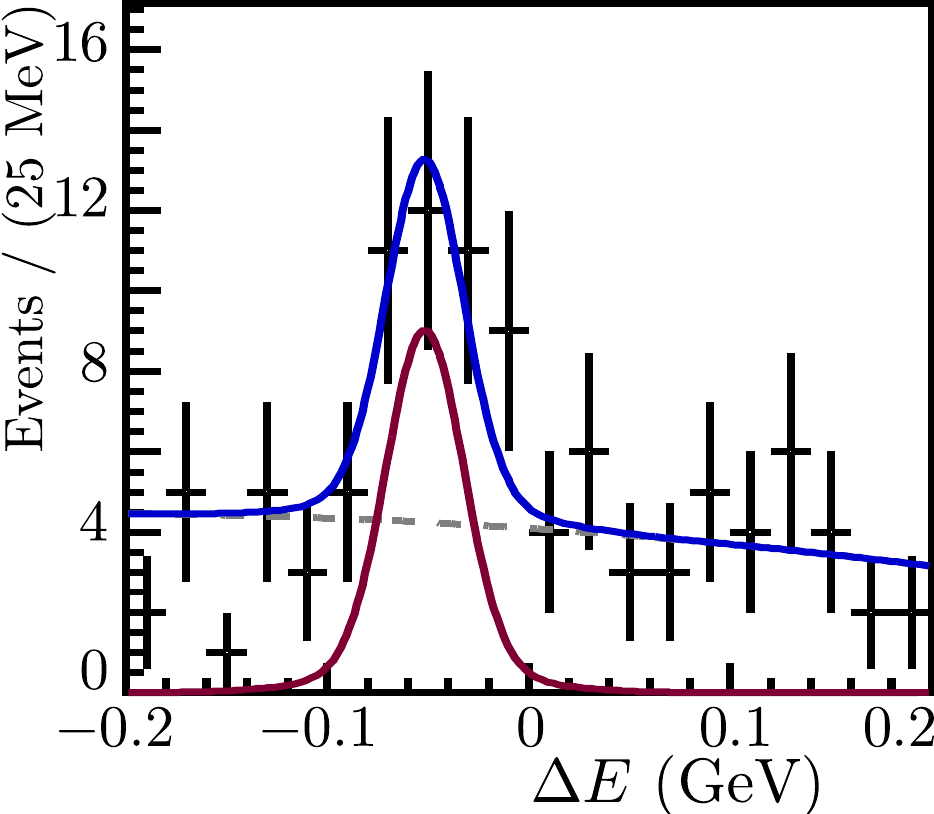}
  \end{minipage}~~~~
  \begin{minipage}{0.36\linewidth}
    \caption{\label{fig:kk}Distributions, similarly to Fig.~1, of the $\bs\to K^+K^-$ candidates
      and the fitted PDF as a solid blue curve.
      The solid-red and the dotted-grey curves represent the signal and the continuum
      component of the PDF, respectively.}
  \end{minipage}
\end{figure}

We do observe a 5.8$\sigma$ excess of $24\pm6$ events in the $\bsst\barbsst$ region for the $\bskk$ mode (Fig.~\ref{fig:kk}).
The BF, $\BR(\bskk)=\bfbstokk$, is derived.
However, no significant signal is seen for the other modes.
Including the systematic uncertainties, we set the following upper limits at 90\% confidence level:
$\BR(\bspipi)$ $<\bfbstopipi$, $\BR(\bskpi)<\bfbstokpi$ and, assuming $\BR(\bs\to K^0\bar K^0)=2\times\BR(\bs\to\KS\KS)$, $\BR(\bskzkz)<\bfbstokzkz$.
The later is the first limit set for the $\bskzkz$ mode.
All the other values are compatible with the CDF results \cite{PRL_97_211802,PRL_103_031801}.

\section{Study of $\pmb{\bsjpsietaP}$ and Search for $\pmb{\bsjpsifz}$}\label{sec:jpsi}
\begin{figure}[!htb]
  \centering
  \begin{minipage}{0.4\linewidth}
    \centering
    \includegraphics[width=\linewidth,height=4cm]{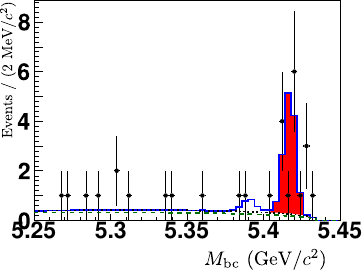}
  \end{minipage}~~~
  \begin{minipage}{0.4\linewidth}
    \centering
    \includegraphics[width=\linewidth,height=4cm]{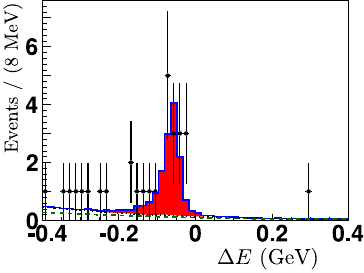}
  \end{minipage}
  \caption{\label{fig:jpsieta}$\mbc$ (left) and $\deltae$ (right) distributions, similarly to Fig.~1,
    of the $\bsjpsieta$ candidates
    (points with error bars) and the fitted PDF as a solid curve.
    The sub-modes $\eta\to\gamma\gamma$ and $\eta\to\pi^+\pi^-\pi^0$, which are fitted separately, are summed in these plots.
    The green-dotted curve (plain red region) represents the continuum (signal) component of the PDF.
    The small peak in the $\mbc$ plot is the signal contribution in the $\bsst\bar\bs$ region.}
\end{figure}

\begin{figure}[!htb]
  \centering
  \begin{minipage}{0.6\linewidth}
    \centering
  \includegraphics[width=\linewidth]{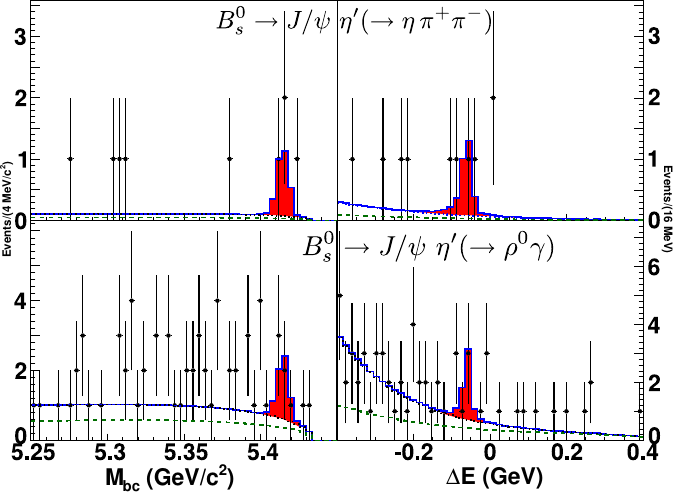}
  \end{minipage}
  ~~
  \begin{minipage}{0.35\linewidth}
    \caption{\label{fig:jpsietap}$\mbc$ (left) and $\deltae$ (right) distributions, similarly to Fig.~1, 
      of the $\bsjpsietap$ candidates
      (points with error bars) and the fitted PDF as a solid curve.
      The green-dotted curve represents the continuum component of the PDF.
      The red region represents the signal component of the PDF.
    }
  \end{minipage}
\end{figure}

Results about the first observation of
$\bsjpsieta$ and the first evidence for $\bsjpsietap$  are reported~\cite{hepex_0912_1434}.
The $\jpsi$ candidates are formed with oppositely-charged electron or muon pairs,
while $\eta$ candidates are reconstructed via the $\eta\to\gamma\gamma$ and $\eta\to\pi^+\pi^-\pi^0$ modes.
A mass (mass and vertex) constrained fit is then applied to the $\eta$ ($\jpsi$) candidates.
The $\eta^{'}$ candidates are reconstructed via the $\eta^{'}\to\eta\,\pi^+\pi^-$ and $\eta^{'}\to\rho^0\gamma$ modes,
while the $\rho^0$ candidates are selected from $\pi^+\pi^-$ pairs.
If more than one candidate per event satisfies all the selection criteria, the one with the smallest fit residual is selected.
The main background is the continuum, which is reduced by requiring $R_2<0.4$. 
The combined $\mbc$ and $\deltae$ distributions are presented in Figs.~\ref{fig:jpsieta} ($\bsjpsieta$) and \ref{fig:jpsietap} ($\bsjpsietap$).
We obtain $\BR(\bsjpsieta)=\bfbstojpsieta$ and $\BR(\bsjpsietap)=\bfbstojpsietaprime$.
This is, respectively, the first observation ($7.3\sigma$) and the first evidence (3.8$\sigma$) for these modes.

The $\bsjpsifz$ mode is especially interesting for the hadron-collider experiments because it has only four charged tracks in its final state.
Recent calculations and measurements predict the ratio $\rfphi=\frac{\BR(\bsjpsifz)\times\BR(\fz\to\pi^+\pi^-)}{\BR(\bs\to\jpsi\,\phi)\times\BR(\phi\to K^+K^-)}$
to be $\approx0.2$ \cite{PRD_79_074024}, $0.42\pm0.11$ \cite{PRD_80_052009} or $\approx0.24$ \cite{PRD_81_074001},
in agreement with other predictions from generalized QCD factorization \cite{PRD_82_076006}.

\begin{figure}[!htb]
  \centering
  \includegraphics[width=0.4\linewidth,height=4cm]{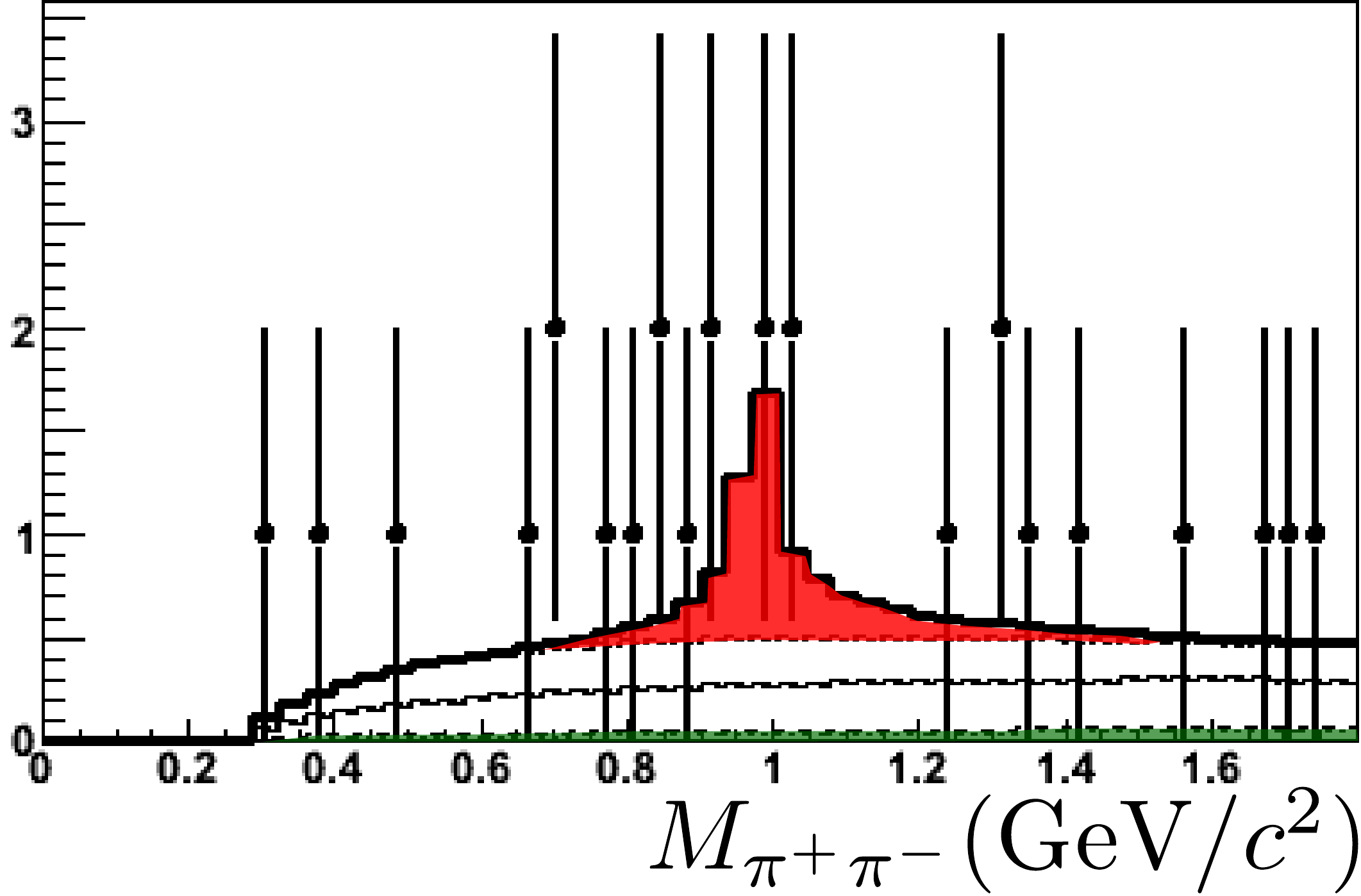}
  ~~
  \includegraphics[width=0.4\linewidth,height=4cm]{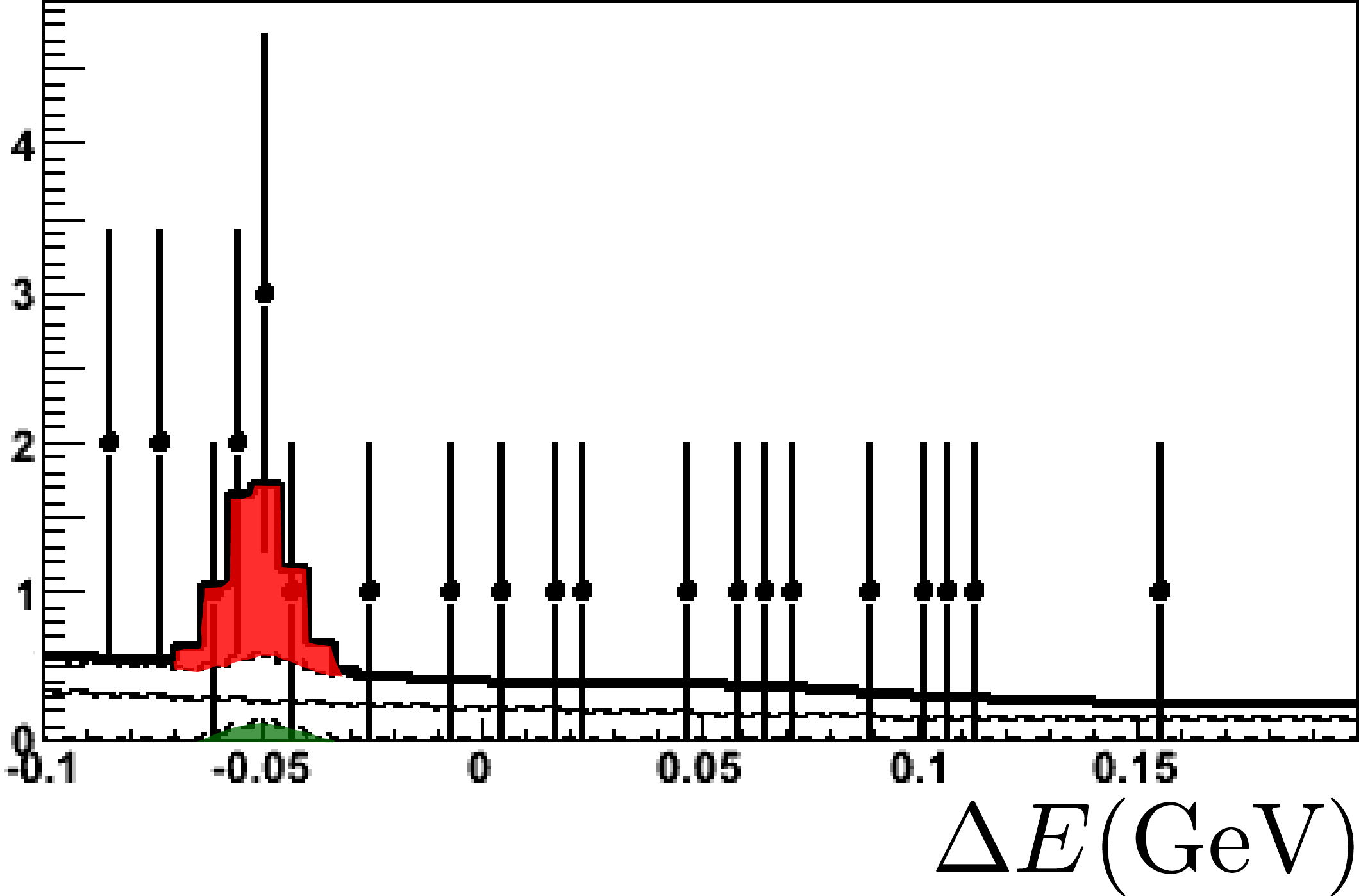}
  \caption{\label{fig:fz} $\fz$ mass (left) and $\deltae$ (right) distributions of the $\bsjpsifz$ candidates.
    The solid-black curve is the total fitted PDF.
    The plain green region represents the contribution of the non-resonant $\bs\to\jpsi\,\pi^+\pi^-$, while the plain red region is the signal.
    The dotted-black curve is the contribution of the other $\bs\to\jpsi\,X$ modes.}
\end{figure}

With the same selection for the $\jpsi$ as described above, and the reconstruction of $\fz\to\pi^+\pi^-$ candidates,
the $\bsjpsifz$ signal is fitted using the distribution of the energy difference, $\deltae$, and the $\fz$ mass, $M_{\pi^+\pi^-}$ (Fig.~\ref{fig:fz}).
No significant signal ($6.0\pm4.4$ events, 1.7$\sigma$) is seen and we set the upper limit \cite{hepex_1009_2605} $$\BR(\bsjpsifz)\times\BR(\fz\to\pi^+\pi^-)<\bfbstojpsifz\,,$$
or, similarly, $\rfphi<0.275$ (at 90\% C.L.) using our preliminary result of $\BR(\bs\to\jpsi\,\phi)$ \cite{hepex_0905_4345}.
These limits are clearly in the region of interest and an update using our full data sample (120$\invfb$) and an improved selection is being performed.

\section{Study of $\pmb{\bs\to\dsSTdsST}$ and Measurement of $\pmb{\dG/\G}$}
\begin{figure}[!htb]
  \centering
  \begin{minipage}{0.65\linewidth}
    \centering
    \includegraphics[width=\linewidth]{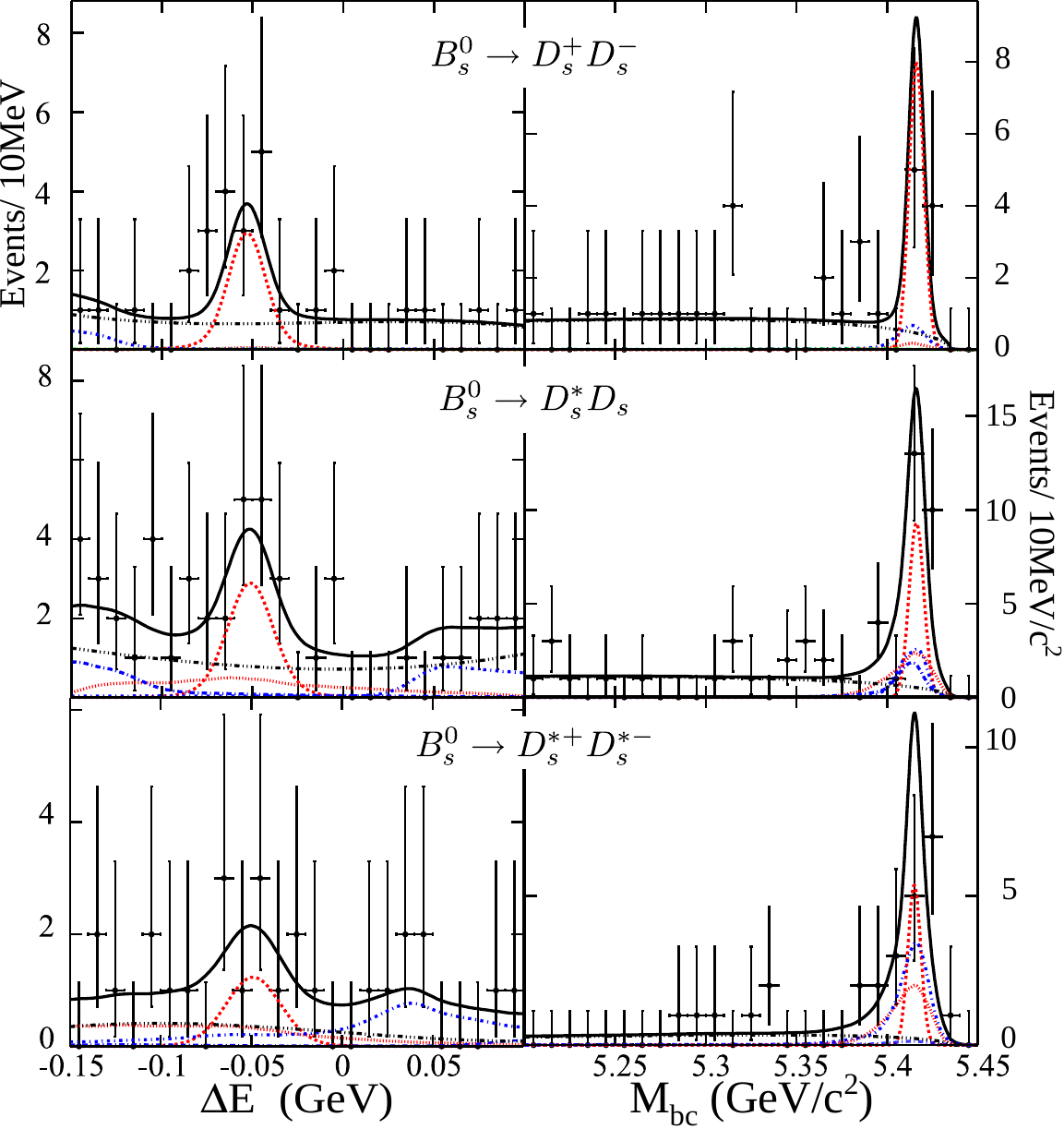}
  \end{minipage}
  ~~
  \begin{minipage}{0.32\linewidth}
    \caption{\label{fig:dsstdsst}$\deltae$ (left) and $\mbc$ (right) distributions, similarly to Fig.~1,
      of the $\bs\to D_s^+D_s^-$ (top), $\bs\to D_s^{\ast\pm}D_s^{\mp}$ (middle) and $\bs\to D_s^{\ast+}D_s^{\ast-}$ (bottom) candidates,
      together with the fitted PDF.
      Except the continuum background component, which is shown by the black dashed-dotted curve, all the other contributions are peaking in $\mbc$.
      The correct (wrong) combination signal, shown by the peaking (smooth) red dashed curve and the cross-feed components, shown by the blue dashed-dotted curve are well separated in $\deltae$.
    }
  \end{minipage}
\end{figure}

We finally report the results from our analysis of the $\bs\to\dsSTdsST$ decays \cite{PRL_105_201802}.
These modes are $CP$ eigenstates \emph{and} CKM favored  ($b\to c\bar cs$ transition of order $\lambda^2$).
In the heavy-quark limit, they are $CP$ even and dominate $\Delta\Gamma$ \cite{PLB_316_567}.
The relative width difference of the $\bs-\bar\bs$ system can be obtained from the relation
$\dG/\G=\frac{2\times\BR(\bs\to\dsSTdsST)}{1-\BR(\bs\to\dsSTdsST)}$.
In order to reconstruct the $\bs\to\dsSTdsST$ candidates, we form $\ds$ candidates from 6 modes:
$\ds\to\phi\pi^-$, $\ds\to K^{\ast0}K^-$, $\ds\to\KS K^-$, $\ds\to\phi\rho^-$, $\ds\to K^{\ast0}K^{\ast-}$ and $\ds\to\KS K^{\ast-}$.
Only one candidate per event is selected based on the values of $M(\ds)$ and $M(\dsst)-M(\ds)$.
The same likelihood as in Sec.~\ref{sec:bshh} is used to reject 80\% of the continuum events, while 95\% of the signal is kept.
The $\deltae$ and $\mbc$ distributions for each of the three $\bs\to\dsSTdsST$ modes are fitted simultaneously.
The signal PDF is made of two components studied with signal MC simulations: 
the correctly reconstructed candidates and the wrong combinations in which a non-signal track (photon) is included in place of a true daughter track (photon).
In addition the so-called cross-feed contributions are included:
a $\dsstds$ ($\dsstdsst$) event can be selected as a $\dsds$ ($\dsstds$) candidate with a lower energy because one photon is missing; 
conversely, a $\dsds$ ($\dsstds$) candidate can be reconstructed as a $\dsstds$ ($\dsstdsst$) candidate with an additional photon, hence its energy larger than expected.

\begin{table}[!tb]
  \centering
  \caption{\label{tab:dsstdsst} Signal event yields, $N_{\rm sig.}$, significances, $S$, including systematics and BFs, $\BR$, for the three $\bs\to\dsSTdsST$ modes and their sum.}
  \begin{tabular}{ccccc}
    \br
    Mode              &$N_{\rm sig.}$      &$S$        &$\BR$             &$\BR$ World Average \cite{PDG10}\\ 
    \mr
    $\bs\to\dsstdsst$ &$4.9^{+1.9}_{-1.7}$ &3.2$\sigma$&$\bfbstodsstdsst$ &First evidence\\
    $\bs\to\dsstds$   &$9.2^{+2.8}_{-2.4}$ &6.6$\sigma$&$\bfbstodsstds$   &First observation\\
    $\bs\to\dsds$     &$8.5^{+3.2}_{-2.6}$ &6.2$\sigma$&$\bfbstodsds$     &$(1.04^{+0.37}_{-0.34})\%$ \\
    \mr
    $\bs\to\dsSTdsST$ &$22.6^{+4.7}_{-3.9}$&           &$\bfbstodsSTdsST$ &$(4.0\pm1.5)\%$\\
    \br
  \end{tabular}
\end{table}

The fit results can be seen in Fig.~\ref{fig:dsstdsst} while the numerical values are reported in Table~\ref{tab:dsstdsst}.
With the relation above, we extract $$\dG/\G=\dGoG\,.$$  
This value is in agreement with the SM expectations \cite{JHEP_06_072} and with the results from ALEPH, $(25^{+21}_{-14})\%$ \cite{PLB_486_286}, 
D\O, $(7.2\pm3.0)\%$ \cite{PRL_102_091801}, and CDF, $(12^{+9}_{-10})\%$ \cite{PRL_100_121803}.
With only 23 fully-reconstructed signal events, our measurement is already competitive with the Tevatron values.

\section{Conclusion}
We presented recent results on $\bs$ decays obtained from 23.6 $\invfb$ of $\FiveS$ data recorded by the Belle detector.
While modes with large statistics can provide precise measurements of BFs and $\bsST$ properties, 
first observations of several $CP$-eigenstate $\bs$ decays are a confirmation
of the large potential of our 120$\invfb$ $e^+e^-\to\FiveS$ data sample and advocate an ambitious $\bs$ program at super-$B$ factories.

\section*{References}
\bibliography{bib}{}

\providecommand{\newblock}{}
\begin{thebibliography}{10}
\expandafter\ifx\csname url\endcsname\relax
  \def\url#1{{\tt #1}}\fi
\expandafter\ifx\csname urlprefix\endcsname\relax\def\urlprefix{URL }\fi
\providecommand{\eprint}[2][]{\url{#2}}
% Bibliography created with iopart-num v2.1
% /biblio/bibtex/contrib/iopart-num

\bibitem{NIMA_479_117}
{A.~Abashian \emph{et al.} (Belle Collaboration)} 2002 {\em
  Nucl.~Instrum.~Methods Phys.~Res., Sect.~A\/} {\bf 479} 117.

\bibitem{NIMA_499_1}
{S.~Kurokawa and E.~Kikutani} 2003 {\em {Nucl.~Instrum.~Methods Phys.~Res.,
  Sect.~A}\/} {\bf 499} 1.

\bibitem{PRD_52_3048}
{I.~Dunietz} 1995 {\em {Phys.~Rev.~D}\/} {\bf 52} 3048;
{I.~Dunietz, R.~Fleischer and U.~Nierste} 2001 {\em {Phys.~Rev.~D}\/} {\bf 63}
  114015.

\bibitem{ZPC_54_653}
{R.~Aleksan, I.~Dunietz and B.~Kayser} 1992 {\em {Z.~Phys.~C}\/} {\bf 54} 653;
{R.~Fleicher} 2003 {\em {Nucl.~Phys.~B}\/} {\bf 671} 459.

\bibitem{PRD_77_054010}
{S.~Nandi and U.~Nierste} 2008 {\em {Phys.~Rev.~D}\/} {\bf 77} 054010.

\bibitem{PRD_70_031502}
{D.~London and J.~Matias} 2004 {\em {Phys.~Rev.~D}\/} {\bf 70} 031502.

\bibitem{PLB_459_306}
{R.~Fleischer} 1999 {\em {Phys.~Lett.~B}\/} {\bf 459} 306.

\bibitem{PLB_475_111}
{P.~Ball and R.~Fleischer} 2000 {\em {Phys.~Lett.~B}\/} {\bf 475} 111.

\bibitem{PLB_316_567}
{R.~Aleksan {\it et al.}} 1993 {\em {Phys.~Lett.~B}\/} {\bf 316} 567.

\bibitem{PRD_81_112003}
{A.~Drutskoy {\it et al.} (Belle Collaboration)} 2010 {\em {Phys.~Rev.~D}\/} {\bf
  81} 112003.

\bibitem{PRL_100_112001}
{K.F.~Chen \emph{et al.} ({Belle Collaboration})} 2008 {\em Phys.~Rev.~Lett.\/}
  {\bf 100} 112001.

\bibitem{PRL_98_052001}
{A.~Drutskoy \emph{et al.} (Belle Collaboration)} 2007 {\em Phys.~Rev.~Lett.\/}
  {\bf 98} 052001.

\bibitem{PRD_75_012002}
{G.S.~Huang {\it et al.} (CLEO Collaboration)} 2007 {\em Phys.~Rev.~D\/} {\bf 75}
  012002.


\bibitem{PDG10}
{K.~Nakamura {\it et al.} (Particle Data Group)} 2010 {\em J.~Phys.~G\/} {\bf 37}
  075021.

\bibitem{hepex_1010_1589}
{D.~Asner {\it et al.} (Heavy Flavor Averaging Group)} 2010 {arXiv:1010.1589 [hep-ex].}

\bibitem{PRL_102_021801}
{R.~Louvot {\it et al.} (Belle Collaboration)} 2009 {\em {Phys.~Rev.~Lett.}\/}
  {\bf 102} 021801;
2010 {\em {Phys.~Rev.~Lett.}\/}
  {\bf 104} 231801.

\bibitem{PLB_318_549}
{A.~Deandrea {\it et al.}} 1993 {\em {Phys.~Lett.~B}\/} {\bf 318} 549;
{R.H.~Li, C.D.~L\"u and H.~Zou} 2008 {\em Phys.~Rev.~D\/} {\bf 78} 014018.

\bibitem{PRL_51_1945}
{L.~Wolfenstein} 1983 {\em Phys.~Rev.~Lett.\/} {\bf 51} 1945.

\bibitem{PRL_10_531}
{N.~Cabibbo} 1963 {\em Phys.~Rev.~Lett.\/} {\bf 10} 531;
{M.~Kobayashi and T.~Maskawa} 1973 {\em Prog.~Theor.~Phys.\/} {\bf 49} 652.

\bibitem{hepex_0912_4179}
{B.~Adeva {\it et al.} (LHCb Collaboration)} 2009 {arXiv:0912.4179v1 [hep-ex].}

\bibitem{PRL_91_171802}
{B.~Aubert {\it et al.} (BaBar Collaboration)} 2003 {\em {Phys.~Rev.~Lett.}\/}
  {\bf 91} 171802.

\bibitem{PRL_91_201801}
{K.F.~Chen {\it et al.} (Belle Collaboration)} 2003 {\em {Phys.~Rev.~Lett.}\/}
  {\bf 91} 201801.

\bibitem{PRL_41_1581}
{G.C.~Fox and S.~Wolfram} 1978 {\em Phys.~Rev.~Lett.\/} {\bf 41} 1581.

\bibitem{PRL_103_191802}
{T.~Aaltonen \emph{et al.} ({CDF Collaboration})} 2009 {\em Phys.~Rev.~Lett.\/}
  {\bf 103} 191802.

\bibitem{PRD_82_072007}
{C.C.~Peng {\it et al.} (Belle Collaboration)} 2010 {\em {Phys.~Rev.~D}\/} {\bf
  82} 072007.


\bibitem{PRL_91_261801}
{S.H.~Lee {\it et al.} (Belle Collaboration)} 2003 {\em {Phys.~Rev.~Lett.}\/} {\bf
  91} 261801.

\bibitem{PRL_97_211802}
{A.~Abulencia {\it et al.} (CDF collaboration)} 2006 {\em {Phys.~Rev.~Lett.}\/}
  {\bf 97} 211802.

\bibitem{PRL_103_031801}
{T.~Aaltonen \emph{et al.} ({CDF Collaboration})} 2009 {\em Phys.~Rev.~Lett.\/}
  {\bf 103} 031801.

\bibitem{hepex_0912_1434}
{I.~Adachi {\it et al.} (Belle Collaboration)} 2009 {arXiv:0912.1434
  [hep-ex]}.

\bibitem{PRD_79_074024}
{S.~Stone and L.~Zhang} 2009 {\em {Phys.~Rev.~D}\/} {\bf 79} 074024.

\bibitem{PRD_80_052009}
{K.M.~Ecklund {\it et al.} (CLEO Collaboration)} 2009 {\em {Phys.~Rev.~D}\/} {\bf
  80} 052009.

\bibitem{PRD_81_074001}
{P.~Colangelo, F.~De Fazio and W.~Wang} 2010 {\em {Phys.~Rev.~D}\/} {\bf 81}
  074001.

\bibitem{PRD_82_076006}
{O.~Leitner {\it et al.}} 2010 {\em {Phys.~Rev.~D}\/} {\bf 82} 076006.

\bibitem{hepex_1009_2605}
{R.~Louvot} 2010 {\it Proceedings of Science} PoS(FPCP 2010)015.

\bibitem{hepex_0905_4345}
{R.~Louvot} 2009 arXiv:0905.4345v2 [hep-ex].

\bibitem{PRL_105_201802}
{S.~Esen {\it et al.} (Belle Collaboration)} 2010 {\em {Phys.~Rev.~Lett.}\/} {\bf
  105} 201802.

\bibitem{JHEP_06_072}
{A.~Lenz and U.~Nierste} 2007 {\em {J.~High Energy Phys.}\/}  {JHEP06(2007)072}.

\bibitem{PLB_486_286}
{R.~Barate {\it et al.} (ALEPH Collaboration)} 2000 {\em {Phys.~Lett.~B}\/} {\bf
  486} 286.

\bibitem{PRL_102_091801}
{V.M.~Abazov \emph{et al.} ({D0 Collaboration})} 2009 {\em Phys.~Rev.~Lett.\/}
  {\bf 102} 091801.

\bibitem{PRL_100_121803}
{T.~Aaltonen \emph{et al.} ({CDF Collaboration})} 2008 {\em Phys.~Rev.~Lett.\/}
  {\bf 100} 121803.

\end{thebibliography}
\bibliographystyle{iopart-num}   

\end{document}